# Liquid droplet morphology on the fiber of a fog harvester mesh and the droplet detachment conditions under gravity

Arani Mukhopadhyay, Partha Sarathi Dutta, Amitava Datta, and Ranjan Ganguly

Department of Power Engineering, Jadavpur University, Kolkata-700098, India

**ABSTRACT**

Liquid droplets on fiber are often observed both in nature and in different engineering applications, like a fog harvesting mesh. Knowledge about drop-on-fiber morphology and its shedding under the influence of gravity can allow for the design of better separation technology. Mutual interaction of surface tension forces arising out of the surface energies of the liquid and the fiber solid, and the weight of the liquid droplet gives rise to different morphologies of the droplet, which may occur in a stable or meta-stable configuration. Predicting the droplet shape on a fiber of specified dimension and surface wettability accurately for a given volume of liquid is challenging since the curvature of both the droplet and the cylinder influence the phenomenon. We have numerically investigated the droplet shape and transition criterion for various volumes at different contact angles under the effect of varying Bond numbers using an open-source surface evolver code. It is observed that depending upon the relative dimensions of the liquid droplet, the fiber diameter, wettability, and gravity, the liquid exists on the fiber either in "barrel" or in "clamshell" shape. A relation between shedding volume and the Bond number is deduced, and the detachment volumes are calculated.

## 1. INTRODUCTION

The accentuating crisis of freshwater in the last few decades has led to a serious drive in technology development toward water sustainability [1]. Harvesting fog from atmospheric [2] and industrial sources [3] have emerged as a viable alternative. In these applications, a fog-laden stream is allowed to pass over a mesh of metal or polypropylene fibers, when the fog droplets deposit on the fibers by inertial impaction, physical interception, and Brownian diffusion [4]. With progressive deposition, the liquid volume on the mesh increases, leading to the formation of droplets of different shapes. These droplets in turn offer enhanced hydrodynamic resistance to the oncoming fog stream and alter the aerodynamic and deposition efficiencies of the mesh. Knowledge of the shape of the deposited droplet is important because the deposited volume of liquid droplet increases the effective shade coefficient (the fraction of the projected area occupied by the mesh fibers plus the adherent liquid on it [5]) of the mesh by partially or completely blocking the mesh pores. Besides, the aerodynamic drag on the deposited droplet, which may often lead to droplet carryover (see Fig 1(a)), is also influenced by the deposited droplet volume and morphology. Finally, as the liquid volume grows in size due to the successive deposition of fog droplets on the mesh fibers, it tends to drip under the influence of gravity or slide down the mesh. It has been shown [5] that premature dripping of the fog water from the mesh is both deterrent and detrimental for the collection [6]. To maximize the collection efficiency of the fog-net, the fog droplets deposited on the fog net must slide down the inclined mesh into the collector without prematurely detaching from the mesh surface due to gravity (see Fig 1(a)). Similar technology in filtration and separation processes that use fibers to separate liquid droplets from a flowing stream is designed to optimize drainage or collection. Therefore, it is very important to know the morphological evolution of a droplet on the mesh fiber ensuing from fog harvesting, and the conditions leading to drip-off due to gravity.

A logical step to understanding the behavior of deposited fog-water on the mesh is to focus on the behavior of a single droplet on an isolated, cylindrical fiber of the mesh. The droplet morphology in such a case would depend on the surface tension of the liquid (water in this case), the deposited liquid volume, the fiber diameter, and its wettability. The shape and stability of a droplet on fiber have posed an intriguing challenge to multiple researches over the past few decades. There are two possible shapes a droplet can assume on a fiber: barrel shape and clamshell shape (see Fig 1(b)) [7]. It is known that for higher liquid volume and lower contact angles, a barrel shape is energetically stable whereas for lower liquid volumes and higher contact angles clamshell shaped droplets are more stable. But in-between such regimes of absolute stability, there exist metastability criteria in which both shapes may emerge as energetically preferred. When a metastable barrel droplet is subjected to non-axisymmetric disturbances, it changes to a clamshell which enhances the chances of its removal. Carroll [8] examined the solution of the Laplace equation for excess pressure in a barrel-shaped droplet and derived the relation among the droplet profile, volume, and stability. McHale et al. [9] later took up this solution along with numerical results for a clamshell shape while using a finite element approach to probe into inflection in droplet profile and stability. It was seen that the stability of the droplet was lost when inflection in the profile of the barrel-shaped droplet had reduced enough to touch the fiber surface. The point of inflection was found near to the contact line of drop and fiber [10]. To understand the relationship between the inflection angle and volume of a drop, 2-D images of a drop were taken and a polynomial fitting method for fiber (PFMF) was used [10]. The location of the boundary was identified with a polynomial fit to measure the contact angle and volume of the drop with an accuracy of the order of 1%.

Factors like global geometry and surface chemistry significantly influence the wetting properties of a surface; for example, decreasing the reduced volume or increasing the contact angle results in a "roll-up" of the barrel to clamshell shape [7]. The dependence of the roll-up process on wettability under negligible gravitational influence was further studied by the application of functionalized fibers for observing these transition volumes [11]. It was seen that under low Bond number ($Bo = \rho g r^2 / \gamma$, where $r$ = the fiber radius, $\rho$ = liquid density, g = acceleration due to gravity and $\gamma$ = surface tension of the liquid) scenario, droplets can exist mainly in three regimes viz., the two absolute-stability regimes (for barrel and clamshell shapes) and the coexistence (or meta-stable) regime wherein both these shapes were independently stable. However certain technological applications require knowledge of droplet shapes when gravity is more pronounced and this scenario was taken up by Chou et al. [12], which showed that the droplet exists primarily in three regimes which include downward clamshell shapes, an independent coexistence regime, and falling off.

The effect of external forces on droplet morphology, like electrowetting, magnetic, gravitational, and drag forces, has also been extensively analyzed. Functional fibers with tunable wettability use electrowetting to change surface energy [11] and have wide applications in filtration and separation technology. The wettability of such functionalized fibers can be reversibly switched. Variable magnetic forces have been used to detach ferrofluid droplets from fibers of varying radii and wettability to arrive at a semi-empirical expression of detachment force for contact angles greater than 20º [13]. For a non-axisymmetric droplet subjected to gravitational effects, the relative size of the droplet formed was found to be inversely proportional to the fiber radius [14]. Hence, filtration technology and functional clothing prefer fine fibers, because it can hold larger droplets. The shape of the contact line of a droplet on fiber is crucial to understanding the effect of gravity on the geometric shape of the droplet and has been estimated to be a 3D spiral line. The swaying of droplets in the Reynolds transition flow regime governs droplet coalescence and growth and has its applications in air filtration [15].

The objective of the present work is to conduct a numerical study to ascertain conditions for the emergence of the competing droplet shapes and observe the transition volumes on a cylindrical, horizontal, smooth fiber under the effects of varying the Bond number (Bo). The study is further extended to evaluate the criterion of gravitation-induced detachment from the fiber. Reduced volumes ($V_R$), defined as the ratio of the liquid volume to the cube of the fiber radius, have been used for all volumetric measurements. The effect of varying the $V_R$ against contact angle for different sets of Bo has been reported. The observation sheds light on the transition from clamshell to barrel metastable state (and vice versa), and the states of absolute stability. The detachment volume when the drop falls from the fiber for a given contact angle and Bo has also been evaluated.

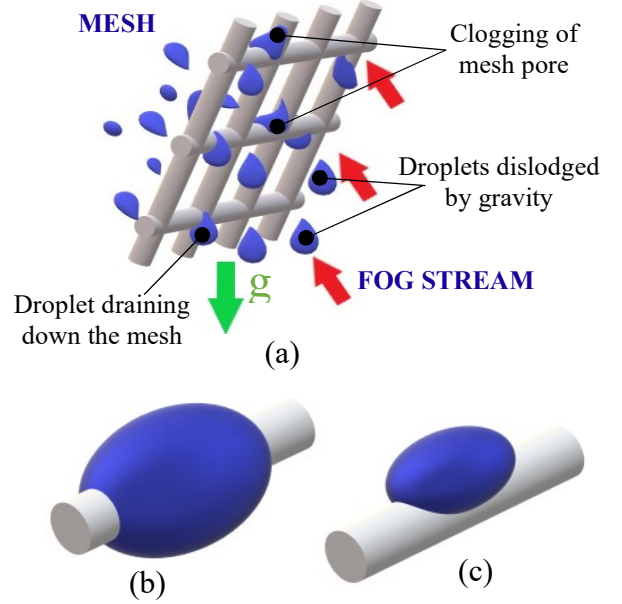

Figure 1: (a) Re-entrainment and premature dripping of liquid droplets from a fog-harvester mesh, affecting collection. (b) Barrel, and (c) Clamshell shapes of liquid on a single fiber.

## 2. THEORETICAL BACKGROUND

It is well known in the field of capillarity that a liquid bounded by a curved surface experiences an excess pressure, which is known as the Laplace pressure. Under conditions involving very low Bo, the Laplace excess pressure is given by Eq. (1) and is uniform throughout the drop. For a droplet on a smooth fiber existing in an axisymmetric shape, i.e. the barrel shape, this equation has been solved using elliptic integrals to determine relevant quantities pertaining to droplet geometry.

$$\Delta p = (p - p_0) = \gamma_{LV}\left(\frac{1}{R_1} + \frac{1}{R_2}\right) \qquad (1)$$

However, no such exact solution exists if the axial symmetry breaks down, in cases when the effect of gravity is considered, or the droplet assumes clamshell shape, or both happens. The minimum volume considered in the study is of the order of 0.512 μL; therefore, the effect of line tension prevalent for nanoscopic droplets can be neglected [16]. Thus, the equilibrium behavior of the free gas-liquid interface of the droplet, expressed implicitly as $F = f(x,y,z)$, on a smooth, solid cylindrical surface, denoted by $S = s(x,y,z)$, is dictated by the Young-Laplace equation [17, 18], which relates the Laplace pressure $p$ (the pressure difference across the liquid interface) with the mean curvature $\kappa_m$ of the interface, the gravity ($g$) and the surface tension ($\gamma$) as

$$2\gamma_{LV}\kappa_m = (p - p_0) - \Delta\rho \vec{g} \cdot \vec{z} \qquad (2)$$

Where $\Delta\rho$ denotes the density difference between the liquid and the gas and $z$ is the local elevation of the surface (from the datum). Goldman [19] showed that for any implicitly defined surface, the mean curvature can be expressed as,

$$\kappa_m = \frac{\nabla F \; Hess(F) \; \nabla F^T - |\nabla F|^2 \; Trace(Hess(F))}{2|\nabla F|^3} \quad (3)$$

Where *Hess(F)* represents the hessian of the surface *F* [19]. Thus, the solution of Eq. (2) can be carried out to deduce the shape of a droplet of specified volume under the following two boundary conditions. The first boundary condition (BC 1) represent the liquid-solid interface on the cylindrically curved, smooth surface of the wire, while the second boundary condition (BC 2) specifies the surface wettability in terms of the contact angle $\theta$ and $\hat{n}_S$ (the normal to the surface S).

$$\text{BC 1:} \; f(S) = 0 \text{ where } S(x,y,z) = 0 \quad (4)$$

$$\text{BC 2:} \; \nabla F . \hat{n}_S = |\nabla F| \sin\theta \text{ on } S(x,y,z) = 0 \quad (5)$$

The problems involve resolving the Laplace equation and the surface morphology in three-dimension for the clamshell geometry, which warrants the use of numerical simulations, e.g., using a standard surface evolver software. Surface Evolver (SE) is an open access finite element package that moves triangulated surfaces in an iterative procedure to minimize a user-defined total energy function for a described geometry using a gradient search algorithm [20]. User input comprises of an initial geometry along with boundary constraints and energy integrals defined over the surface or volume. Simulations are started with an arbitrary geometry (e.g., prismatic) shape of liquid volume on the cylindrical wire. Choice of the initial disposition of this liquid volume (e.g., encompassing the cylinder, akin to the barrel morphology, or one-sided, akin to the clamshell morphology) has been made carefully to ensure convergence. With the passage of iterations, the shape of the liquid volume evolves to minimize the total energy (surface + gravitational potential). Each iteration towards the solution moves through intermediate surface morphologies which do not necessarily correlate with the physical shape of the droplet, and at the end, the shape where no significant decrease of energy after subsequent iteration is observed is considered as the final morphology [21].

To validate the working of our developed numerical model with existing literature, simulations were executed in SE, to evaluate droplet morphology for a reduced volume of 37.8 and contact angle of 30°, under no gravitational influence. These results were previously achieved by McHale et al. [22] and a comparison of the reduced maximal thickness (the ratio of the maximum distance of the droplet from the fiber to the fiber radius) and the reduced wetted length (defined as the ratio of the maximum length of the fiber wetted by the droplet to the fiber radius) for both models has been depicted in Fig. 2. Throughout this study, the mesh edge length ($l_e$) was chosen to have an optimal range of *1.6×10⁻⁴m ≥ $l_e$ ≥ 5.33×10⁻⁵m*. Any edges that were greater or smaller than these limits were either refined or deleted, respectively, during the adaptive grid refinement by the SE algorithm. These values were chosen by trial and error, following similar codes for large droplet simulations used by Berthier and Brakke [23]. A stricter refinement criterion of *1.2×10⁻⁴m ≥ $l_e$ ≥ 4×10⁻⁵m* affected droplet $V_R$ during detachment by only 0.77 percent.

To undertake this comparison, axial and side views of the developed geometry were transferred to CAD software (Fusion 360) and the reduced lengths were measured consequently. A deviation of 0.8% for reduced maximal thickness and 4.3% for reduced wetted length were noted with the pre-existing results. These simulations assumed negligible gravitational effects, which is a typical assumption for an oil droplet on a fiber when completely submerged in water.

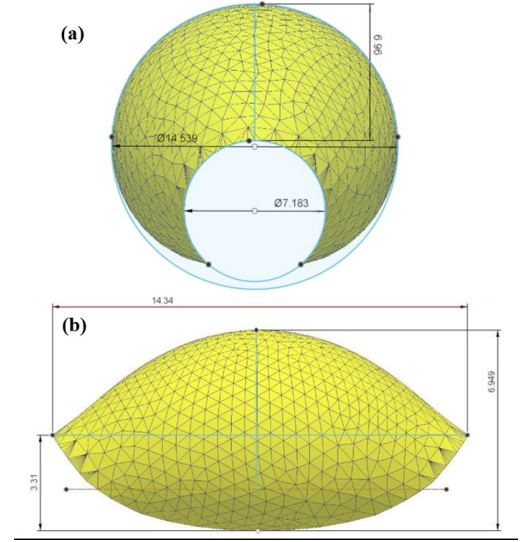

**Figure 2:** Measurement of maximum reduced thickness and reduced wetted length in (a) axial view and (b) side view respectively, for a clamshell droplet, developed in Surface Evolver, in CAD.

For such cases, the total free energy of a drop is mainly dependent on the surface energies and is described by Eq. (6).

$$F = \gamma_{LV} A_{LV} + (\gamma_{SL} - \gamma_{SV}) A_{SL} \quad (6)$$

However, in cases where the effect of gravitation cannot be neglected, i.e. in cases of higher Bond numbers, Eq (6) has to be modified and a volume integral indicating the presence of gravitational forces has to be included. Using Young's equation to relate to the contact angle, we can rewrite Eq. (6) as Eq. (7).

$$F = \gamma_{LV}(A_{LV} - A_{SL}\cos\theta) + \rho g \iiint_V z \; dV \quad (7)$$

As is shown in the later sections, this gravitational effect makes a downward clamshell a preferred choice over a barrel in most cases and a stable barrel shape becomes increasingly rare during greater Bond numbers.

## 3. METHODOLOGY

Having validated the simulation for a simple droplet-on-a-fiber configuration, parametric variations are carried out to identify the conditions of the morphology transition and droplet detachment. For all simulations, smooth horizontal cylindrical fibers have been considered, while distilled water ($\gamma$ = 72 mN/m) has been used for the simulation. The following general methodology has been adopted for the simulations.

### 3.1 Droplet morphology transition criterion

A barrel-shaped droplet exists at low contact angles; thus, to capture the morphology transition phenomenon from a barrel to a clamshell shape, a low intrinsic contact angle (θ, in the range of 10º to 20º) was initially specified to ensure that initially, a barrel-shaped droplet was formed at a given reduced volume. From this initial value, θ was continually increased in steps of 1º in subsequent runs to observe a situation when the converged shape of the liquid morphs into a clamshell shape. This methodology was repeated for different reduced volumes ($V_R$); the resulting plot of the threshold θ (beyond which the clamshell shape was observed) against the corresponding $V_R$ describes the absolute stability regime for the clamshell shape (see the red curve in Fig. 3), which is obtained for a *Bo* of 0.0002.

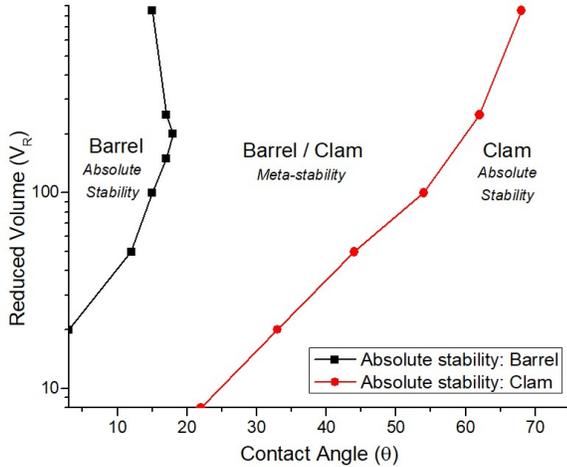

**Figure 3: Comparison of $V_R$ and *θ* where morphology transition occurs in a droplet at *Bo* of 0.0002. Regimes marked 'Barrel' or 'Clamshell' indicate regions of absolute stability, while both shapes are independently stable in the meta-stable region.**

Similarly, to observe the transition from a clamshell shape to a barrel, we began with a droplet in an initial clamshell configuration (θ above the absolute clamshell stability threshold) and kept reducing the contact angle in steps of 1º until the converged solution yielded a barrel shape. The region in the left of the ensuing transition $V_R$ vs θ (the black line) plot in Fig. 3 denotes the absolute stability regime for barrel shape. The simulations at various $V_R$ thus resulted in three stability regimes for each *Bo*. As is shown in Fig. 3 for a *Bo* of 0.0002, the three regimes include a stable clamshell shape and a stable barrel shape interspaced with one metastable state in which both states are independently stable.

Each of these simulations involved the use of multiple iterations which ensured that no changes in non-dimensional total energy in the magnitude of $10^{-12}$ were observed after subsequent iterations before moving to a different contact angle. This ensured that each simulation was complete before changing the contact angle. In cases that involved no gravitational effect (i.e., *Bo*=0), a droplet in a stable barrel state can grow indefinitely large and remain in the barrel shape for any contact angle. However, such would be impossible in a real scenario owing to the presence of even the slightest perturbation forces which help in transitioning the barrels to clamshells. Hence, for the development of Fig. 3, very low gravitational effects (with *Bo* in the range of 0.0002) had to be assumed to ensure that barrel droplets would not grow indefinitely large but would shift to clams at higher contact angles.

### 3.2 Droplet detachment criterion

To ascertain the critical detachment volume, simulations were carried out with an initial volume in the clamshell regime chosen to be small enough so that the drop did not detach from the fiber. This volume was considered as the starting point and the reduced volume in each run was incremented by $\Delta V_R =1.0$ for the given fiber diameter and Bond number. Simulation for a given run was considered complete only when the reduction in non-dimensional total energy between two iterations fell below $10^{-12}$ as compared to the previous iteration. The simulation was terminated when the droplet completely fell off the fiber, i.e., there were no remaining attachments of the drop to the fiber. Different morphologies of a droplet before reaching critical detachment volume have been shown in Fig. 4. In Fig 4, run 12 having a $V_R$ of 3425 would be considered critical ($V_{R,critical}$) for a droplet subjected to normal gravitational constants having a *Bo* of 0.0044 on a surface with a contact angle of 30º.

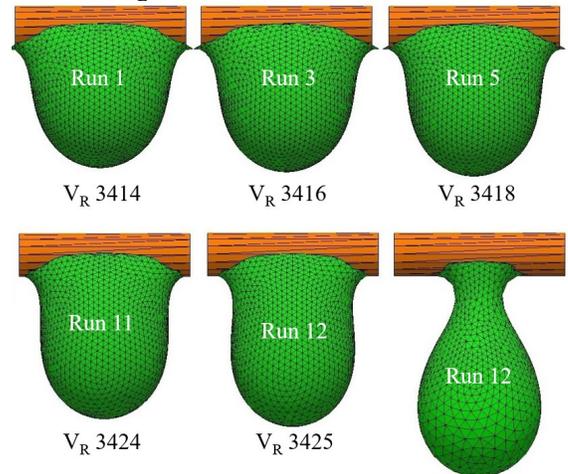

**Figure 4: Growth and detachment simulations for a droplet at *Bo* of 0.0044. Detachment volume ($V_{R,critical}$) is found at iteration 12 of the SE program at a $V_R$ of 3425 for a *θ* of 30º on a fiber of radius $4\times10^{-4}$m.**

However, for the estimation of $V_{R,critical}$ for small θ (< 20º) at lower *Bo* (≤ 0.0044) a different approach had to be adapted. Simulations with an initial clamshell configuration resulted in the transition to a barrel-shaped droplet after a few iterations, owing to the insufficient gravitational forces to make the downward clamshell an energetically preferred choice. Hence, the simulations had to be started in a barrel-shaped configuration, in which the subsequent increase of volume resulted in the barrel shape to ultimately transition into a clamshell, followed by the consequent detachment of the drop from the fiber.

## 4. EFFECT OF THE BOND NUMBER ON MORPHOLOGY AND DETACHMENT

As explained in Section 3, the $V_R$ values at which transition occurs from clamshell shape to barrel shape were calculated at different *Bo*. Fig. 5 shows the variation of the absolute stability in the barrel regime. It is seen that with an increase in the value of *Bo*, the absolute stability regime in the barrel shape, shifts leftward on the $V_R$ - θ plane, i.e., towards lowers $\theta$, indicating that greater surface tension forces are required. For greater $Bo$ ($\geq 0.0044$) this regime is only present at very low $\theta$ ($\leq 5°$). Both barrels and clams are energetically preferred in the meta-stable state which lies on the other side of the absolute stability in the barrel regime. It may be pointed out here, that in the absence of any gravitational effect ($Bo = 0$), barrels can grow indefinitely large. However, even small gravitational effects or perturbations are detrimental to this shape and can make it shift to a clamshell shape when energetically preferred.

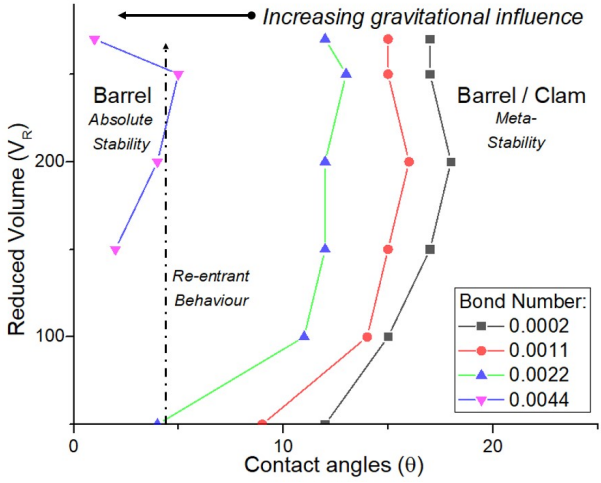

**Figure 5: A comparison of $V_R$ and $\theta$ in the absolute stability in barrel regime, for varying $Bo$. Gravitational influence for each curve on the left is greater than its counterpart on the right.**

Similarly, Fig. 6 shows the variation of the absolute stability regime of the clamshell shaped droplets with varying $Bo$. As gravitational influence becomes more pronounced i.e., in the case of increasing $Bo$, this regime is seen to move towards lower contact angles. This indicates that clamshells become increasingly preferred energetically.

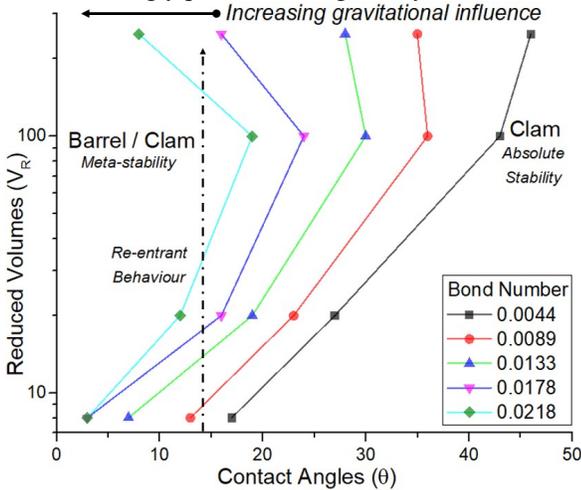

**Figure 6: A comparison of $V_R$ and $\theta$ in the absolute stability in the clamshell regime, for varying $Bo$. Gravitational influence for each curve on the left is greater than its counterpart on the right.**

The dotted lines in Fig. 5 and Fig. 6 demarcate the regimes that demonstrate re-entrant behavior, which was observed earlier by Eral et. al [11]. This behavior is observed during droplet growth on a surface having considerably low contact angles ($\theta$) at higher $Bo$. If we consider a case of droplet growth, at $Bo = 0.0218$ at $\theta \approx 15°$, we can see from Fig. 6 (by following the dotted line) that a droplet exhibiting absolutely stable clamshell shape for $V_R < 35$ shifts to the meta-stable regime for a higher value of $V_R$, where it can also take a barrel shape. On further growth of the liquid, corresponding to $130 < V_R$, it re-enters the absolute stability regime and returns to a clamshell morphology. Continued growth after a clamshell shaped droplet has undergone such re-entrant behavior leads to the droplet becoming unstable and being detached from the fiber.

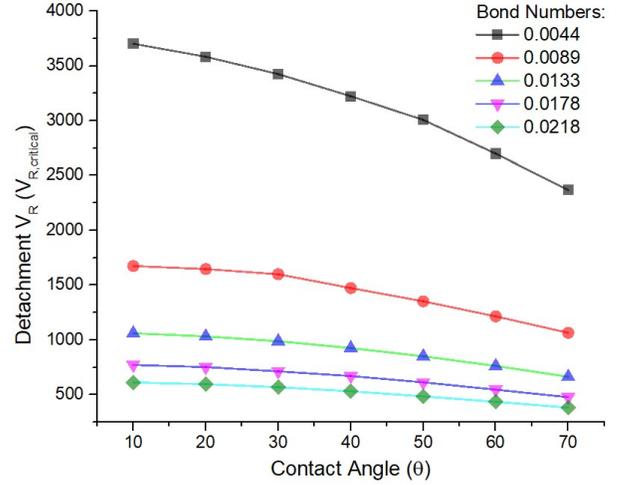

**Figure 7: $V_R$ at which detachment occurs at various contact angles for varying Bond numbers.**

Shedding of a droplet occurs when the droplet volume is sufficiently large and the gravitational forces acting on it can overcome the surface tension forces. Under such conditions, if a droplet is in a barrel shape, it will move into a clamshell configuration before subsequent detachment. Fig. 7 shows the variation of the critical reduced volume, $V_{R,\,critical}$ at which detachment occurs, as functions of $\theta$ under different values of $Bo$. It was observed that the reduced volumes at which detachment occurs for a constant contact angle vary directly with the contact angle ($\theta$) and inversely with $Bo$ following a curve-fit given by Eq. (8), where $A$ is the proportionality constant.

$$V_{R,critical} = A(\cos\theta)^{0.442}(Bo)^{-1.129} \qquad (8)$$

## 5. CONCLUSIONS

Numerical simulations, using an open-source surface evolver software is carried out to study the morphology of liquid droplets on a horizontally placed, smooth cylindrical fiber of varying wettability. The droplets are found to exist in a cylindrical (barrel) shape or a clamshell shape, depending upon the liquid volume, its contact angle with the solid surface, and the Bond number. The reduced volumes at which transition occurs from barrel shape to clamshell shape and vice-versa is calculated under varying gravitational influence. It is seen that with increasing gravitational effects i.e., with increasing Bond number, absolutely stable barrel-shaped droplets become unstable and a pendant clamshell shape is favored. This hints at the instability of the barrel shape to non-axial perturbations or force fields. While it is seen that appreciable stable barrel regime is observed for *Bo* in the range of 0.0044 or smaller, clamshells become increasingly predominant from this point. Although a large portion of the regimes exhibits a metastable state, it is intuitively understood that for greater *Bo* the shift from pendant clamshell to barrel in this state is not spontaneous, and can only happen due to external perturbations or non-axial forces. For higher *Bo*, both in the case of stable clamshell and barrel regimes, re-entrant behavior can be observed.

Besides, the conditions for droplet detachment from the fiber are investigated under different reduced volumes, contact angles, and Bond numbers. It was seen that the critical detachment volume decreases with the increase of *Bo*. Such a relationship holds importance in separation or filtration technology like fog harvesting where efficient drainage ensures lesser clogging and better deposition, thus resulting in better fog collection. Future scope includes the employment of PFMF for estimation of detachment volumes or estimating detachment from fibers of different orientations. Also, dynamic models, taking into account the fog stream velocity and the thermodynamic conditions of the ambient, may be taken up as a future exercise.

## ACKNOWLEDGEMENTS

The authors gratefully acknowledge the funding from DST SERB (Grant No: CRG/2019/005887). AM also thankfully acknowledge the assistantship from JU-RUSA.

## NOMENCLATURE

| Symbol | Description | Units |
|---|---|---|
| $\Delta P$ | Laplace excess pressure | [N/m²] |
| $R_1, R_2$ | Principal radii of curvature | [m] |
| $\gamma_{LV}, \gamma_{SL}, \gamma_{SV}$ | Surface tension at Liquid-Vapor, Solid-Liquid, and Solid-Vapor interface | [N/m] |
| $A_{LV}, A_{SL}, A_{SV}$ | Area at Liquid-Vapor, Solid-Liquid, Solid-Vapor interface | [m²] |
| $\rho$ | Density | [kg/m³] |
| $g$ | Gravitational constant | [m/s²] |
| $z$ | Elemental volume height | [m] |
| $V_R$ | Reduced volume ($V/r_F^3$) | -- |
| $Bo$ | Bond number | -- |